\documentclass[%
5p,hyperref,sort&compress,
 reprint,
 amsmath,amssymb,
 aps,
]{revtex4-2}
\usepackage[
pdfauthor={derajan},
pdftitle={How to do this},
pdfstartview=XYZ,
bookmarks=true,
colorlinks=true,
linkcolor=blue,
urlcolor=blue,
citecolor=blue,
pdftex,
bookmarks=true,
linktocpage=true,   
hyperindex=true
]{hyperref}
\usepackage {braket}
\usepackage{graphicx}
\usepackage{dcolumn}
\usepackage{float}
\usepackage{bm}

\begin{document}

\preprint{APS/123-QED}

\title{Quantum hacking against discrete-modulated continuous-variable quantum key distribution using modified local oscillator intensity attack with random fluctuations}

\author{Lu Fan, Yiming Bian, Mingze Wu} \author{Yichen Zhang} \email{zhangyc@bupt.edu.cn} 
\author{Song Yu}
\affiliation{%
State Key Laboratory of Information Photonics and Optical Communications, School of Electronic Engineering, Beijing University of Posts and Telecommunications, Beijing {100876}, China
}%
\date{\today}
\begin{abstract}
The local oscillator in practical continuous-variable quantum key distribution system fluctuates at any time during the key distribution process, which may open security loopholes for the eavesdropper to hide her eavesdropping behaviors. Based on this, we investigate a more stealthy quantum attack where the eavesdroppers simulates random fluctuations of local oscillator intensity in a practical discrete-modulated continuous-variable quantum key distribution system. Theoretical simulations show that both communicating parties will misestimate channel parameters and overestimate the secret key rate due to the modified attack model, even though they have monitored the mean local oscillator intensity and shot-noise as commonly used. Specifically, the eavesdropper's manipulation of random fluctuations in LO intensity disturbs the parameter estimation in realistic discrete-modulated continuous-variable quantum key distribution system, where the experimental parameters are always used for constraints of the semidefinite program modeling. The modified attack introduced by random fluctuations of local oscillator can only be eliminated by monitoring the local oscillator intensity in real time which places a higher demand on the accuracy of monitoring technology. Moreover, similar quantum hacking will also occur in practical local local oscillator system by manipulating the random fluctuations in pilot intensity, which shows the strong adaptability and the important role of the proposed attack. 
\end{abstract}

\maketitle

\section{\label{sec:level1}Introduction}
Quantum key distribution (QKD) \cite{BB84} relies on quantum mechanical properties to ensure communication security, which allows two remote parties to generate unconditionally secure keys despite the presence of potential eavesdroppers in quantum channels \cite{Pirandola:20, RevModPhys.92.025002}. Depending on the dimension of coding, it can be divided into discrete-variable QKD (DV-QKD) and continuous-variable QKD (CV-QKD) \cite{RevModPhys.84.621, e17096072}. Between them, CV-QKD uses coherent detection technology to replace single-photon detection technology, which can be well compatible with classical optical communication systems \cite{toward_2021}. In addition, the transmission distance achieved recently can support the requirement in metropolitan distances \cite{toward_2018, 50, wang2022sub} which facilitates the large-scale application of QKD \cite{jouguet2013experimental,zhang2019integrated,18.3,zhang2020long,huang2021experimental,zhang2023experimental,PTMP}.

CV-QKD based on Gaussian-modulation (GM) \cite{PhysRevLett.88.057902, PhysRevLett.93.170504, navascues2006optimality, garcia2006unconditional,pirandola2008characterization, li2014continuous,zhang2014continuous,pirandola2015high, tian2022experimental,jain2022practical,pi2023sub} is widely used, where the quadratures of quantum states are modulated according to the Gaussian probability distribution.
Moreover, discrete-modulated (DM) CV-QKD \cite{leverrier2009unconditional}, where the quadratures of quantum states are modulated discretely, has also been gaining a lot of attention. Due to the simplicity of state preparation \cite{wang2022sub,li2018user, matsuura2021finite} and lower complexity of classical error correction \cite{pan2022experimental,roumestan2022experimental}, the unconditional security of DM CV-QKD protocols (e.g.two-state and three-state protocols \cite{zhao2009asymptotic, bradler2018security}) has been successively proved. Recently, one of its security analysis theory is using the semidefinite program (SDP) model to give the lower bound of the secret key rate \cite{ghorai2019asymptotic,lin2019asymptotic,lin2020trusted,denys2021explicit, liu2021homodyne}, and related experiments have also been carried out \cite{wang2022sub, pan2022experimental,pereira2022probabilistic}.
With the tremendous progress in theory and experiment of DM CV-QKD in recent years, its practical system security has received increasing attention.
Different from theoretical analysis, the transmission of the local oscillator in the practical DM CV-QKD system would leave Eve with vulnerabilities to perform quantum hacking.

The local oscillator (LO) is required for stable coherent detection, whose non-ideality within practical environments is usually ignored in the security analysis of CV-QKD. Due to the imperfection of practical blocks and devices, researchers have proposed many quantum hacking models aiming at practical LO, such as local oscillator calibration attack \cite{jouguet2013preventing}, local oscillator intensity fluctuation attack \cite{PhysRevA.88.022339}, wavelength
attacks \cite{huang2013quantum, ma2013wavelength}, polarization attack \cite{zhao2018polarization} and so on. Particularly, the transmission of LO opens a security vulnerability which gives eavesdropper (Eve) advantages to exploit the fluctuation of LO intensity \cite{ma2014enhancement}. We note that in the previous studies \cite{PhysRevA.88.022339}, Eve is assumed to reduce the intensity of LO overall using variable attenuators which can easily be defended by commonly-used monitoring technology such as monitoring the LO intensity mean and optical power or monitoring the shot-noise. In other words, the above manipulation and monitoring of LO intensity is carried out on its overall mean. 
When Eve simulates random fluctuations in LO intensity with invariant mean value, a more serious security loophole appears because above mentioned monitoring techniques will fail.
\begin{figure*}[t]
\includegraphics[width=16cm]{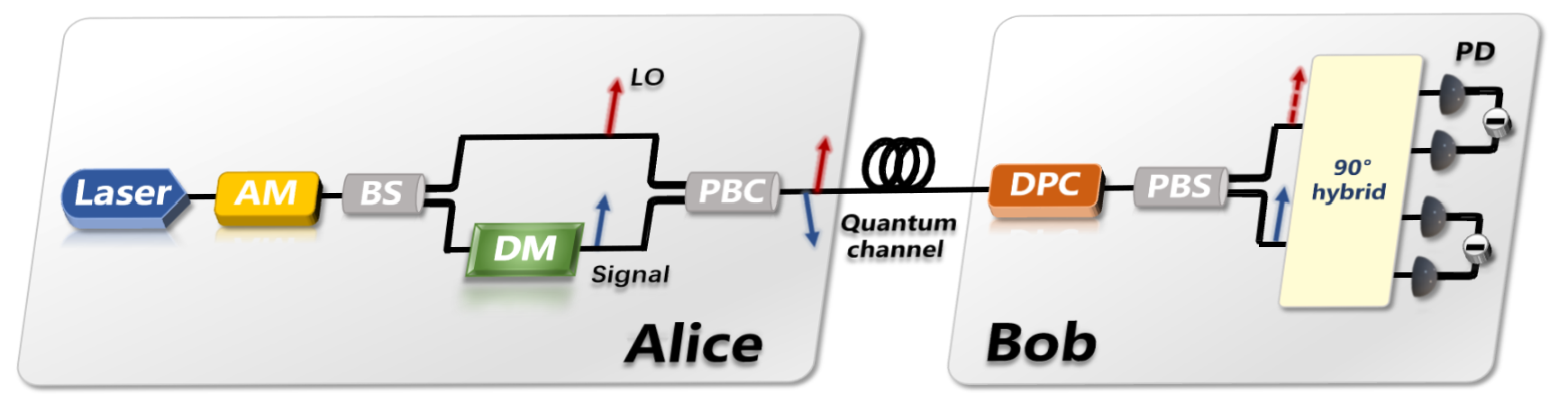}
\caption{\label{fig0}The practical block diagram of DM CV-QKD system. AM, amplitude modulation; BS, beam splitter; PBC, polarization beam coupler; DPC, dynamic polarization controller; PBS, polarizing beam splitter; PD, photodiodes.}
\end{figure*}

\begin{figure*}[t]
\includegraphics[width=18cm]{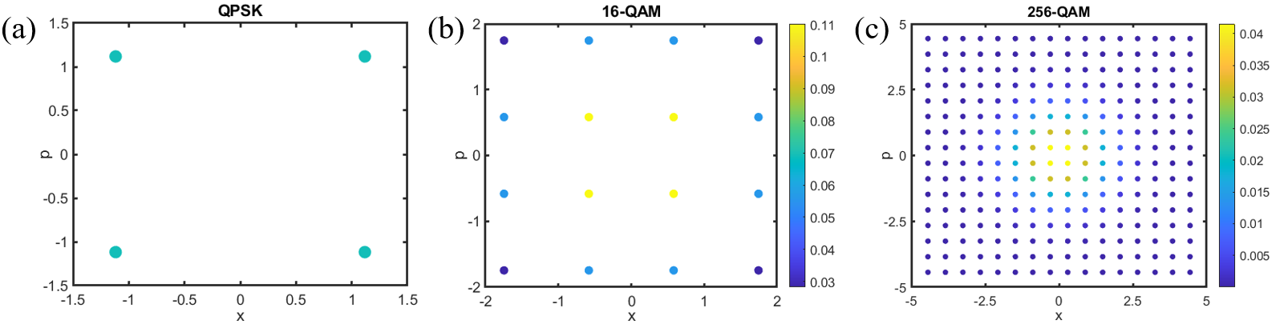}
\caption{\label{figccc} The constellation plots of different discrete modulation schemes are presented. The regular components of the coherent state are no longer continuous, but discrete. The first corresponds to the constellation plot of quadrature phase shift keying (QPSK). The latter two correspond to the constellation plots of 16 quadrature amplitude modulation (16-QAM) with $\nu=0.085$ and 256 quadrature amplitude modulation (256-QAM) with $\nu=0.035$ where $\nu$ is the Gaussian-like distribution parameter. Colors indicate the probabilities corresponding to each coherent state.}
\end{figure*}

In this paper, we propose a modified local oscillator intensity attack (LOIA) with random fluctuations.
Eve can exploit the random fluctuating properties of LO intensity to launch a more stealthy attack by attenuating a part of LO pulses while amplifying another part of LO pulses.
The security of the practical DM CV-QKD system under this attack is considered, which mainly includes the effect of random fluctuations in LO intensity on the estimation of experimental parameters in SDP security analysis model. Quantitative theoretical simulations support that Eve's manipulation of random fluctuations in LO intensity would cause insecure final keys to be shared between legitimate parties. 
Remarkably, the fluctuating variance of LO intensity is founded to be the core character and even no secure key could be generated when the fluctuating variance is relatively large. 
While seriously affecting system security, the modified attack model has the advantage of evading traditional commonly-used monitoring techniques.
Therefore, the communication parties must employ real-time LO intensity monitoring with higher accuracy to evade such the attack. The study complements the research of quantum hacking of the practical DM CV-QKD system.

This paper is organized as follows. In Section II, we first introduce the modified LOIA with random fluctuations, and then we perform the practical security analysis of DM CV-QKD under the modified attack model in Section III. Theoretical simulations prove the feasibility of the attack in Section IV. The discussion is shown in Section V and finally the conclusion is given in Section VI.

\section{\label{sec2:level1}The modified LOIA with random fluctuations}
The practical DM CV-QKD system mainly includes optical source, discrete modulation, quantum channel, heterodyne detector and so on, as shown in FIG. \ref{fig0}. 
The coherent light generated by the laser is amplitude modulated to obtain optical pulses that meet the requirements of the certain extinction ratio and duty cycle, and then Alice divides them into two parts using the optical BS, one part includes LO pulses with strong power and the other part includes the signal pulses with weak power. Furthermore, Alice encodes the key information on the signal pulses through discrete modulation, where the prepared coherent state is selected from a finite number of constellations in phase space.
Time multiplexing technology and polarization multiplexing technology allow modulated signal pulses and LO pulses to be transmitted in the same channel without interference in the scheme.
Bob demultiplexes the received pulses and then obtains the outcomes using the heterodyne detector. 

\begin{figure*}[t]
\includegraphics[width=16cm]{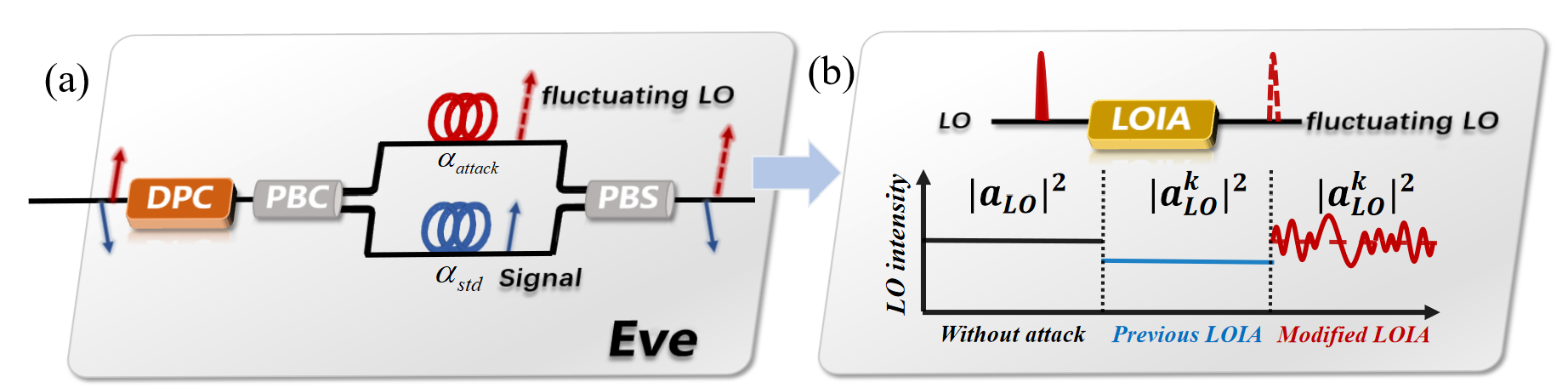}
\caption{\label{attack}$(a)$The implementation of the modified LOIA with random fluctuations. Due to Eve's manipulation, the LO and signal pulses pass through fibers with different attenuation factors. The LO received at the receiver end fluctuates randomly around the mean value. $(b)$The details of Eve's quantum hacking attack on LO. The line at the left side (black) refers to initially calibrated LO intensity; the line in the middle (blue) is the instantaneous LO intensity under the previous LOIA model where Eve attenuates the LO intensity overall; the line at the right side (red) is the instantaneous LO intensity under the modified LOIA model. Eve simulates the random fluctuations of LO intensity to launch the LOIA.}
\end{figure*}
As can be seen from the constellation plot of two different kinds of discrete modulation (i.e. QPSK and QAM) in FIG. \ref{figccc}, each coherent state represents a constellation. In phase space, the regular components $x$ and $p$ of the coherent state are discretely distributed.  
For the more commonly-used QAM modulation method, the mutual information decreases compared with the Shannon capacity when its constellations are uniformly distributed. Probability constellation shaping (PCS) was therefore introduced to address this limitation, with the idea of using a non-uniform distribution on a QAM lattice. The probability distribution for such a PCS M-QAM is
\begin{equation}\label{eq:ganssian}
p_{x,p}\sim\exp(-\nu(x^2+p^2)).
\end{equation}
This distribution depends on $\nu\textgreater0$ and the spacing between the possible values of $x$ (or $p$). The parameter, $\nu$ allows changing the shape of the Gaussian-like distribution on the constellation, which can be optimized to maximize the key rate. 

Before the setup of such a CV-QKD system, the communication parties will calibrate the LO intensity value in advance to obtain the shot-noise unit (SNU). The transmission process of LO is always ideally assumed safe in the theoretical security analysis while there are some imperfections in the practical system. During the transmission of LO pulses on the quantum channel, Eve simulates the random fluctuations in LO intensity around its mean value to cover up her attack on signal pulses that successfully steals final keys, which can be called the LOIA with random fluctuation.
The modified attack LOIA with random fluctuations can be carried out in many ways where one of the implementations is presented in FIG. \ref{attack}$(a)$. 
Eve may actively open a security loophole: She intercepts the signal pulses and the LO pulses, replacing the imperfect channel between the communication parties as her own quantum channel. The LO pulses are transmitted on the attacked channel with attenuation factor $\alpha_{attack}$ and the signal pulse are transmitted on the normal fiber channel with attenuation factor $\alpha_{std}$. Without changing the phase of LO, Eve can manipulate LO intensity to simulate random fluctuations by changing the attenuation factor $\alpha_{attack}$. Eve then couples the fluctuating LO pulses and signal pulses together and sends them to the receiver. 
The attack effect is shown in  FIG. \ref{attack}$(b)$. 
The calibrated SNU in advance is defined as $u_S=A^2a_{LO}^2$ where LO is considered immutable during the key distribution process.
Here $A$ represents the amplification within the heterodyne detector and $I_{LO}=a_{LO}^2$ represents the intensity of LO. However, LO intensity fluctuates randomly around its mean value at any time during the key distribution process due to Eve's manipulation. Whereas the previous LOIA model considered the overall attenuation of LO, here Eve carries out the modified LOIA model where the fluctuation in LO intensity is a random variable with invariant mean to evade the LO intensity mean monitoring and shot-noise monitoring. The accurate description of LO intensity fluctuations will be directly related to the upper limit of the amount of final keys that Eve could steal. Therefore, the practical LO intensity $\vert{a_{LO}^{k}}\vert^2=k{\vert{a_{LO}\vert}^2}$ where the attack factor $k$ is introduced to indicate the random fluctuation of LO intensity. To simplify the analysis, we can describe the attack model through some following conditions:

$(a)$ $k$ is an independent random variable with the same distribution;

$(b)$ $k$ follows some probability distributions with a mean of ${E_k}={1}$ and variance ${V_k}$. 

$(c)$ $k$ is independent of the LO intensity ${I_{LO}}$.

$(d)$ The probability distribution function (PDF) of $k$ will not change during CV-QKD process.

The practical SNU in the presence of the attack will be $u_S^k=ku_S$ which is different from the previous calibrated SNU.
If Alice and Bob still use the calibrated SNU for parameter estimation, the system's excess noise will be misestimated and insecure secret keys will be generated.
As a result, this attack model reveals a more serious loophole that cannot be avoided by commonly-used monitoring techniques.
\section{\label{sec:level1}The MODIFIED LOIA against PRACTICAL DM CV-QKD system}
Unlike the particularity of Gaussian modulation, the covariance of Alice and Bob for discrete modulation with finite constellations cannot be constructed, so there is no way to directly obtain corresponding covariance matrix that conforms to entanglement-based (EB) model. Recent studies have shown that the security analysis of DM CV-QKD can be performed using the SDP modeling, which represents the calculation of secret key rate as a convex optimization problem. The difference of SNU in the modified LOIA from the calibration SNU affects the normalization of the experimental parameters used for constraints in the SDP modeling.
\subsection{\label{sec2:level1}The SDP modeling of DM CV-QKD protocol}
In the prepare-and-measure (PM) version of the DM CV-QKD protocol, a set of coherent states $\{\ket{\alpha_k}\}$ are prepared, where Alice chooses one of the state $\ket{\alpha_k}$ to be sent with the probability $p_k$. The average state sent by Alice can be represented by the density matrix $\tau$ given by the weighted mixture of coherent states,
\begin{equation}\label{eq:tao}
\tau:=\sum_k{p_k\ket{\alpha_k}\bra{\alpha_k}}.
\end{equation}
We define the quadrature operators $\hat{x}$ and $\hat{p}$ in phase space as $\hat{x}:=\hat{a}+\hat{a}^{\dagger}$ and $\hat{p}:=-i(\hat{b}+\hat{b}^{\dagger})$ where $\hat{a}$ and $\hat{a}^{\dagger}$ (resp. $\hat{b}$ and $\hat{b}^{\dagger}$) are the annihilation and generation operators of Alice's register $A$ (resp. Bob's register $B$) on Fock space. 
In the EB version of the DM CV-QKD protocol, Alice prepares the bipartite state $\ket{\Phi_{AA'}}$ and makes local POVM measurements of the reserved mode $A$,
\begin{equation}\label{eq:phi}
\ket{\Phi_{AA'}}=\sum_{k=1}^{M}\sqrt{p_k}\ket{\psi_k}\ket{\alpha_k}
\end{equation}
where the $\{\ket{\psi_k}\}$ form an orthonormal basis and $M$ stands for the number of constellations.
Then another mode $A'$ to be sent collapses to the corresponding coherent state. Alice sends $A'$ through the quantum channel $\mathcal{N}_{A'\rightarrow{B}}$ and Bob obtains the received mode $B$. The quantum state shared by communication parties is denoted as $\rho_{AB}=
(id_A\otimes\mathcal{N}_{A'\rightarrow{B}})
(\ket{\Phi_{AA'}}\bra{\Phi_{AA'}})$ where $id_A$ stands for the identity operator acting on mode $A$.

When Eve performs the collective attack, the secret key rate for reverse reconciliation can be calculated using the Devetak-Winter formula,
\begin{equation}\label{eq:keyrate0}
R=\beta I_{AB}-\mathop{sup}\limits_{\mathcal{N}_{A'\rightarrow{B}}}\ {\chi}_{BE},
\end{equation}
where $\beta$ is the reconciliation efficiency \cite{van2004reconciliation}.
$I_{AB}$ is the mutual information between Alice and Bob, and $\chi_{BE}$ is the Holevo bound which decribes the mutual information between Bob and Eve.
The covariance matrix between Alice and Bob is assumed to be in general form,
\begin{align}\label{eq:CovMat1}
\gamma_{AB}=
\begin{pmatrix}
  V\cdot I_2 & Z\cdot\sigma_z \\
Z\cdot\sigma_z & W\cdot I_2 
\end{pmatrix}
,
\end{align}
where $I_2=
\begin{pmatrix}
  1 & 0 \\
  0 & 1
\end{pmatrix}$
,
$\sigma_z=
\begin{pmatrix}
  1 & 0 \\
  0 & -1
\end{pmatrix}$
. The parameter $V$, $Z$, $W$ can be given by
$V:=1+2tr(\rho{\hat{a}}^{\dagger}\hat{a})$, $Z:=tr(\rho({\hat{a}}{\hat{b}}+{\hat{a}}^{\dagger}{\hat{b}}^{\dagger}))$ and $W:=1+2tr(\rho{\hat{b}}^{\dagger}\hat{b})$.
The Holevo bound can be given by
\begin{equation}\label{eq:Holevo}
\chi_{BE}=\sum_{i=1}^{2}G(\frac{\lambda_i-1}{2})-G(\frac{\lambda_3-1}{2}),
\end{equation}
$\lambda_{1\sim2}$ represent the symplectic eigenvalues of $\gamma_{AB}$, and $\lambda_{3}=V-\frac{{Z}^2}{1+W}$. Both $V$ and $W$ can be observed by Alice and Bob locally, there is left unknown $Z:=tr(\rho{C})$ with $C=\hat{a}\hat{b}+\hat{a}^{\dagger}{\hat{b}}^{\dagger}$ in the covariance matrix, which needs to be bounded using some constraints to obtain the secret key rate.

In the security analysis method of SDP modeling, the objective function is the minimum value of the covariance $Z:=tr(\rho{C})$.
The first constraint merely says that $\rho_{AB}$ is obtained by applying the channel $\mathcal{N}_{A'\rightarrow{B}}$ to $\ket{\Phi_{AA'}}$:
\begin{equation}\label{eq:SDP1}
tr_B(\rho_{AB})=tr_B((id_A\otimes\mathcal{N}_{A'\rightarrow{B}})(\ket{\Phi_{AA'}}\bra{\Phi_{AA'}}))=\bar{\tau}
\end{equation}
where we define $\bar{\tau}$ to be the complex conjugate of $\tau$ in the Fock basis.
The other constraints come from observations from the PM scheme: the second moment constraint and two first moment constraints.
The second moment constraint is that defining the operator $\Pi\otimes{\hat{b}^{\dagger}\hat{b}}$ 
where $\Pi:=\sum_k{\ket{\psi_k}\bra{\psi_k}}$ is a projector and observing that
\begin{equation}\label{eq:SDP2}
tr(\rho(\Pi\otimes{\hat{b}^{\dagger}\hat{b}}))=n_B,
\end{equation}
where $n_B$ can be measured in the protocol. 
An operator is introduced to define the first moment constraints,
\begin{equation}\label{eq:tau}
\alpha_{\tau}:=\tau^{1/2}\alpha\tau^{-1/2}.
\end{equation}
Then the constraints are obtained:
\begin{equation}\label{eq:SDP3}
tr(\rho{C_1})=2c_1,
\end{equation}
\begin{equation}\label{eq:SDP4}
tr(\rho{C_2})=2c_2.
\end{equation}
The operators ${C_1}$ and ${C_2}$ is defined by
\begin{equation}\label{eq:C1}
C_1:=\sum_k\overline{\bra{\alpha_k}\alpha_{\tau}\ket{\alpha_k}}\ket{\psi_k}\bra{\psi_k}\otimes{\hat{b}}+h.c.,
\end{equation}
\begin{equation}\label{eq:C2}
C_2:=\sum_k{\bar\alpha_k}\ket{\psi_k}\bra{\psi_k}\otimes{\hat{b}}+h.c..
 \end{equation}
The correlation coefficients $c_1$ and $c_2$ can be estimated experimentally, and $h.c.$ stands for Hermitian conjugate. The last constraint is that the quantum state is semidefinite. Then the SDP can be listed \cite{denys2021explicit},
$min\ tr(\rho C)$,
 \begin{equation}
s.t. 
\left \{
\begin{array}{lr}
    tr_B(\rho)=\bar{\tau} ,\\
tr(\rho(\Pi\otimes{\hat{b^{\dagger}}\hat{b}}))=n_B ,\\
tr(\rho{\sum_k\overline{\bra{\alpha_k}\alpha_{\tau}\ket{\alpha_k}}\ket{\psi_k}\bra{\psi_k}\otimes{\hat{b}}+h.c.})=2c_1 ,\\
tr(\rho{\sum_k{\bar\alpha_k}\ket{\psi_k}\bra{\psi_k}\otimes{\hat{b}}+h.c.})=2c_2, \\
    \rho\succeq0 .
\end{array}
\right.
\end{equation}

Based on the SDP modeling, the analytical lower bound of $Z$ can be also obtained directly as 
\begin{equation}
Z^*:=2c_1-2((n_B-\frac{c_2^2}{\left\langle{n}\right\rangle})w)^{1/2},
\end{equation}
where the average photo number $\left\langle{n}\right\rangle$ and  the quantity $w$ is defined by the protocol
\begin{equation}\label{eq:photot number}
\left\langle{n}\right\rangle:=\sum_k^M{p_k|\alpha_k|^2}
\end{equation}
\begin{equation}\label{eq:w}
w:=\sum_k^M{p_k(\bra{\alpha_k}a_{\tau}^{\dagger}a_{\tau}\ket{\alpha_k}-|\bra{\alpha_k}a_{\tau}\ket{\alpha_k}|^2)}.
\end{equation}
The quantities $c_1$, $c_2$ and $n_B$ are estimated from experiments whose values are related to SNU. 
Therefore, the impact of our proposed attack on SNU will also affect the parameter estimation of $c_1$, $c_2$ and $n_B$ in DM CV-QKD.
\subsection{\label{sec2:level1}The parameter estimation against the modified LOIA with random fluctuations}
It is assumed that $N$ symbols of the original key are used for parameter estimation. The estimation of the three parameters is expressed separately by estimators $\hat{c}_1$, $\hat{c}_2$ and $\hat{n}_B$ \cite{roumestan2022advanced},
\begin{equation}\label{eq:c1}
\hat{c}_1=\frac{1}{\sqrt{u_S}}\frac{\sqrt{2}}{N}\sum_{k=1}^{N}a_ky_k,
\end{equation}
\begin{equation}\label{eq:c2}
\hat{c}_2=\frac{1}{\sqrt{u_S}}\frac{\sqrt{2}}{2N}\sum_{k=1}^{N}x_ky_k,
\end{equation}
\begin{equation}\label{eq:nB}
\hat{n}_B+1=\frac{1}{\sqrt{u_S}}\frac{1}{N}\sum_{k=1}^{N}|y_k|^2,
\end{equation}
where $a_{2k-1}=Re(\bra{\alpha_k}\alpha_{\tau}\ket{\alpha_k})$, $a_{2k}=Im(\bra{\alpha_k}\alpha_{\tau}\ket{\alpha_k})$, $x_k$ is the transmitted symbols with variance $V_A$ from Alice, $y_k$ is the measurement of Bob.
The three parameters $c_1$, $c_2$ and $n_B$ can be inferred from the expected value of estimators, and there are such relationships:
\begin{equation}\label{eq:c12}
{c_1+\frac{3}{4N}c_1}=E\begin{bmatrix}\hat{c}_1\end{bmatrix}=E\begin{bmatrix}\frac{1}{\sqrt{u_S}}\end{bmatrix}E\begin{bmatrix}\frac{\sqrt{2}}{N}\sum_{k=1}^{N}a_ky_k\end{bmatrix},
\end{equation}
\begin{equation}\label{eq:c22}
{c_2+\frac{3}{4N}c_2}=E\begin{bmatrix}\hat{c}_2\end{bmatrix}=E\begin{bmatrix}\frac{1}{\sqrt{u_S}}\end{bmatrix}E\begin{bmatrix}\frac{\sqrt{2}}{2N}\sum_{k=1}^{N}x_ky_k\end{bmatrix},
\end{equation}
\begin{equation}\label{eq:nB2}
{\frac{N}{N-2}(n_B+1)}=E\begin{bmatrix}\hat{n}_B+1\end{bmatrix}=E\begin{bmatrix}\frac{1}{\sqrt{u_S}}\end{bmatrix}E\begin{bmatrix}\frac{1}{N}\sum_{k=1}^{N}|y_k|^2\end{bmatrix}.
\end{equation}

Suppose Eve carries the proposed attack and randomly fluctuates LO intensity, the practical parameters $c_1^k$, $c_2^k$ and $n_B^k$ should be
\begin{equation}\label{eq:c12}
{c_1^k+\frac{3}{4N}c_1^k}=E\begin{bmatrix}\frac{1}{\sqrt{ku_S}}\end{bmatrix}E\begin{bmatrix}\frac{\sqrt{2}}{N}\sum_{k=1}^{N}a_ky_k\end{bmatrix},
\end{equation}
\begin{equation}\label{eq:c22}
{c_2^k+\frac{3}{4N}c_2^k}=E\begin{bmatrix}\frac{1}{\sqrt{ku_S}}\end{bmatrix}E\begin{bmatrix}\frac{\sqrt{2}}{2N}\sum_{k=1}^{N}x_ky_k\end{bmatrix},
\end{equation}
\begin{equation}\label{eq:nB2}
{\frac{N}{N-2}(n_B^k+1)}=E\begin{bmatrix}\frac{1}{\sqrt{ku_S}}\end{bmatrix}E\begin{bmatrix}\frac{1}{N}\sum_{k=1}^{N}|y_k|^2\end{bmatrix}.
\end{equation}
When the communicating parties perform the parameter estimation without the SNU being calibrated again, the estimated value $c_1$, $c_2$ and $n_B$ will be different from the practical value $c_1^k$, $c_2^k$ and $n_B^k$,
\begin{equation}\label{eq:c13}
c_1^k=E[\frac{1}{\sqrt{k}}]c_1\approx{(1+\frac{3}{8}V_k)c_1},
\end{equation}
\begin{equation}\label{eq:c3}
c_2^k=E[\frac{1}{\sqrt{k}}]c_2\approx{(1+\frac{3}{8}V_k)c_2},
\end{equation}
\begin{equation}\label{eq:nB2}
n_B^k+1=E[\frac{1}{{k}}](n_B+1)\approx{(1+V_k)(n_B+1)}.
\end{equation}
From the above analysis, it can be seen that Eve could exploit the random fluctuation in LO intensity to cause incorrect parameter estimation between the communicating parties.
\section{\label{sec:level1}Simulations and results}
To observe the performance of practical DM CV-QKD system under the proposed attack more clearly, the quantum channel between Alice and Bob is modeled as a phase-insensitive Gaussian channel characterized by transmittance $T_c$ and excess noise $\xi_c$.
Against such a Gaussian channel, the practical parameters $c_1^k$, $c_2^k$ and $n_B^k$ can be described as
\begin{equation}\label{eq:C1}
c_1^k=\sqrt{T_c}tr(\bar{\tau}^{-1/2}a\bar{\tau}^{1/2}{a^{\dagger}}),
\end{equation}
\begin{equation}\label{eq:C2}
c_2^k=\sqrt{T_c}\left\langle{n}\right\rangle,
\end{equation}
\begin{equation}\label{eq:C3}
n_B^k+1={T_c}\left\langle{n}\right\rangle+{T_c}{{\xi_c}/2}+1.
\end{equation}
When the communicating parties perform incorrect parameter estimation, the equivalent estimated channel parameters $T_e$ and $\xi_e$ will be:
\begin{equation}\label{te}
T_e=\frac{1}{(1+\frac{3}{8}V_k)^2}{T_c},
\end{equation}
\begin{equation}\label{eq:c3}
\xi_e=\frac{(1+\frac{3}{8}V_k)^2}{1+V_k}\xi_c-(1-\frac{(1+\frac{3}{8}V_k)^2}{1+V_k})V_A-(1-\frac{1}{1+V_k})\frac{2}{T_e}.
\end{equation}
It can be easily seen that the estimated excess noise $\xi_e$ of Gaussian channel is lower than the practical excess noise $\xi_c$ on small random fluctuations of LO intensity, which means that there is a certain probability that the key rate is overestimated, resulting in the security of final secret keys not being guaranteed.
The numerical simulations under the Gaussian channel assumption will be given next.
\begin{figure}[t]
\includegraphics[width=8.6cm]{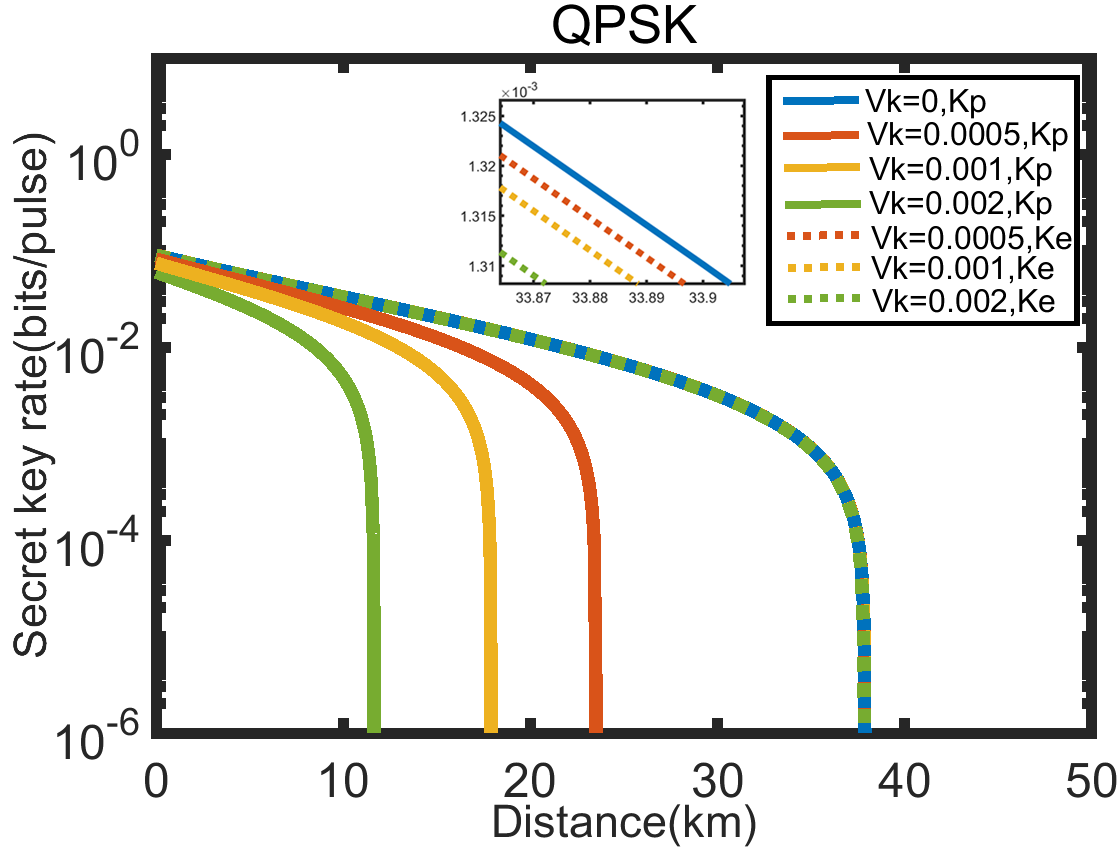}
\caption{\label{QPSK_gaussian}The estimated and practical secret key rate versus transmission distance for the QPSK-modulated protocol against the attack with different fluctuation variances. The modulation variance $V_A=0.456$, the estimated channel excess noise $\xi_e=0.007$ and $\beta=0.95$.}
\end{figure}

FIG. \ref{QPSK_gaussian} compares the estimated key rate $K_e$ and the practical key rate $K_p$ of QPSK-modulated CV-QKD protocol under the modified LOIA with random fluctuation. The modulation variance $V_A=0.456$, the estimated channel excess noise $\xi_e=0.007$ \cite{wang2022sub} and $\beta=0.95$ \cite{ma2021practical}. The solid lines are the secret key rates estimated by Alice and Bob without monitoring the instantaneous LO intensity while the dashed lines reveal the practical secret key rates. Curves of different colors correspond to different degrees of fluctuation in LO intensity with the fluctuation variance $V_k=0, 5\times10^{-4}, 10^{-3}, 2\times10^{-3}$. It can be seen that the practical maximum transmission distance will decrease dramatically if the variances is $2\times10^{-3}$. However, the estimated key rate obtained by both parties through parameter estimation is still close to the key rate without attack. Therefore, this attack will significantly cause both parties to overestimate the secret key rate. 
\begin{figure}[t]
\includegraphics[width=8.6cm]{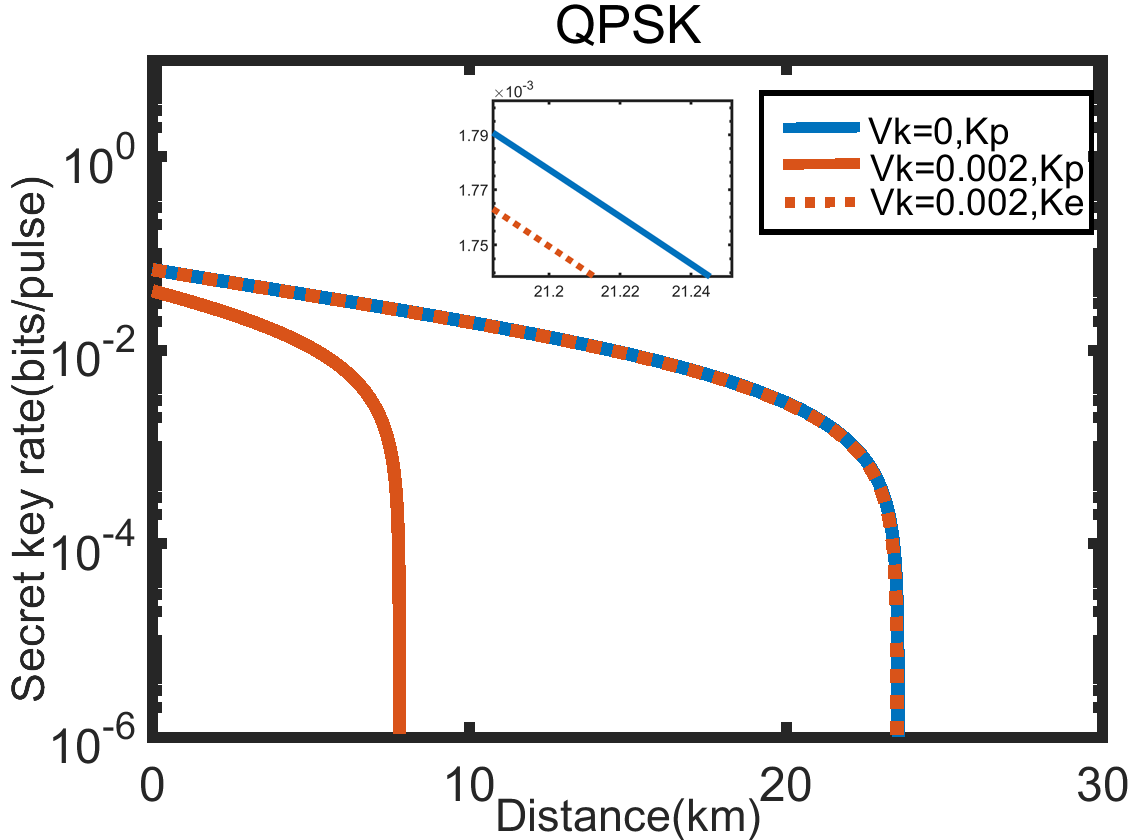}
\caption{\label{QPSK_2}The estimated and practical secret key rate for the QPSK-modulated protocol against the attack with the fluctuation variance $ V_k=2\times10^{-3}$. The modulation variance $V_A=0.456$, the estimated channel excess noise $\xi_e=0.01$ and $\beta=0.95$.}
\end{figure}

FIG. \ref{QPSK_2} clearly shows the amount of overestimated secret key rate by communication parties due to Eve's modified LOIA with the fluctuation variance $V_k=2\times10^{-3}$ and the estimated channel excess noise $\xi_e=0.01$.
The practical maximum transmission distance is reduced to less than $10$ $km$, while the maximum transmission distance estimated by both communicating parties can still reach more than $20$ $km$. The secret keys generated over a communication distance of more than $10$ $km$ are insecure.
It is worth noting that it can be seen from equation \ref{te} that the communicating parties underestimate the transmittance due to the LO intensity random fluctuations.
Therefore, the estimated secret key rate is slightly lower than the ideal one without attack as shown in the simulation plots.
\begin{figure}[t]
\includegraphics[width=8.6cm]{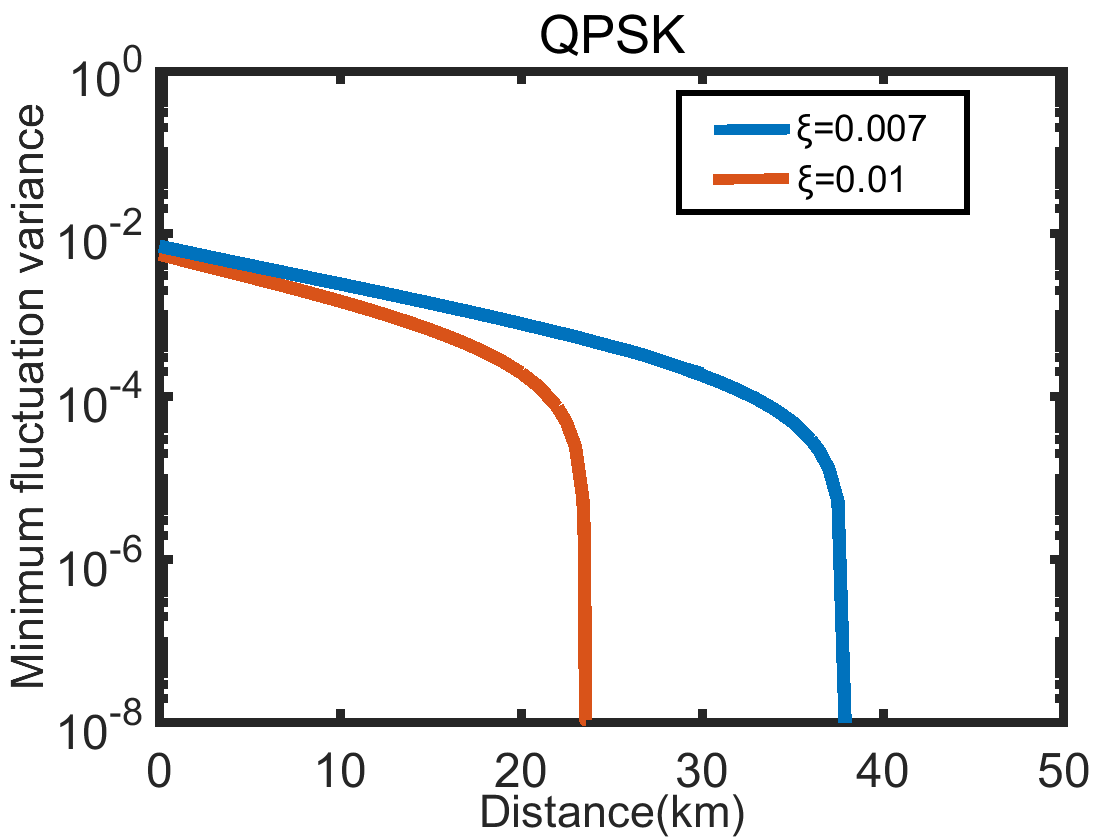}
\caption{\label{QPSK_Vk}The minimal fluctuation variance of LO intensity for the QPSK-modulated protocol where Eve could acquire all keys without being detected. The modulation variance $V_A=0.456$ and the $\beta=0.95$. }
\end{figure}

FIG. \ref{QPSK_Vk} shows the minimum fluctuation variance when Eve could steal all the keys without being detected. Two different solid lines correspond the case with the estimated channel excess noise $\xi_e=0.007$ and $\xi_e=0.01$, respectively. The minimum fluctuation variance is maintained at a low value (lower than $10^{-2}$) which indicates even small fluctuation around initial calibrated LO intensity will severely affect the security. 
Once the fluctuation variance of LO intensity exceeds the minimum value, no secret key can be generated. It can be also seen that as the distance increases or the excess noise increases, the minimum fluctuation variance is getting smaller and the attack effect is better. 
\begin{figure}[t]
\includegraphics[width=8.6cm]{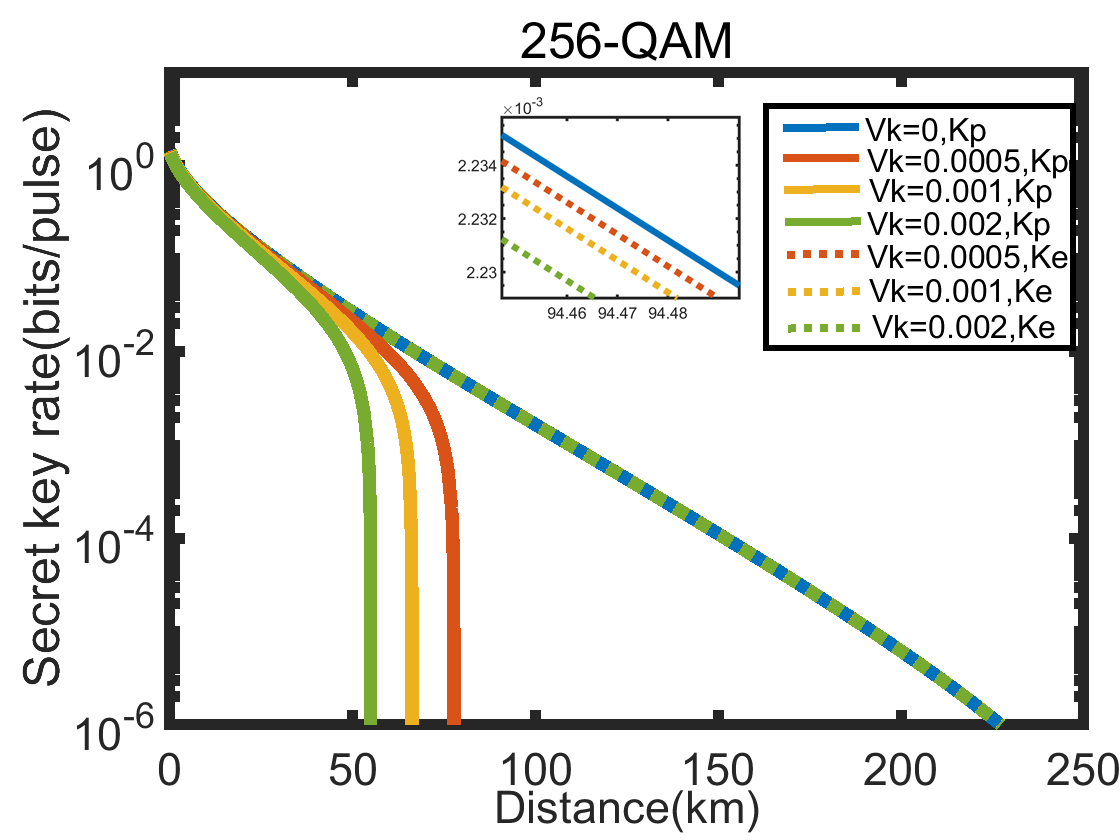}
\caption{\label{tu2}The estimated and practical secret key rate versus transmission distance for the 256-QAM-modulated protocol against the attack with different fluctuation variances. The modulation variance $V_A=6.332$, $\nu=0.039$, the estimated channel excess noise $\xi_e=0.029$ and $\beta=0.95$.}
\end{figure}
\begin{figure}[t]
\includegraphics[width=8.6cm]{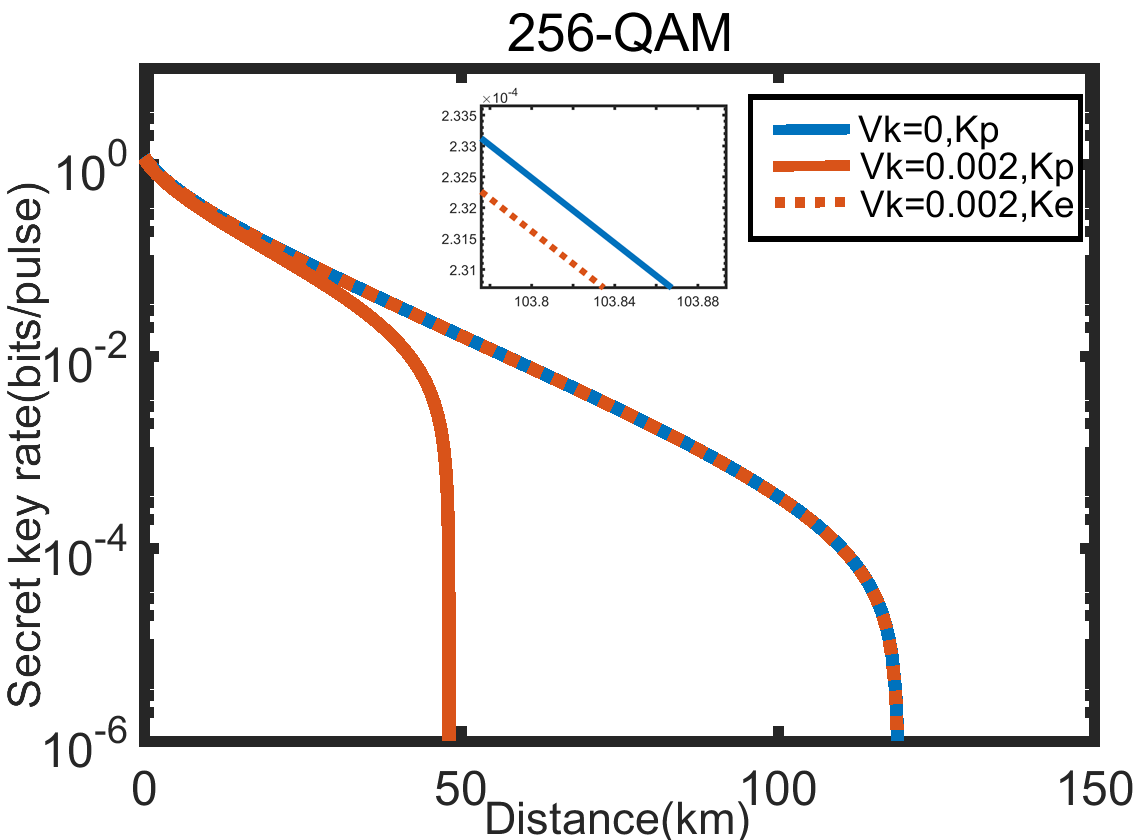}
\caption{\label{QAM_2}The estimated and practical secret key rate for the 256-QAM-modulated protocol against the attack with the fluctuation variance $ V_k=2\times10^{-3}$. The modulation variance $V_A=6.332$, $\nu=0.039$, the estimated channel excess noise $\xi_e=0.05$ and $\beta=0.95$.}
\end{figure}

FIG. \ref{tu2} and FIG. \ref{QAM_2} show the performance under the 256-QAM-modulated protocol against the modified LOIA with random fluctuations. The modulation variance $V_A=6.332$, $\nu=0.039$ \cite{pan2022experimental} and $\beta=0.95$. Similar to QPSK, the fluctuation variance of LO intensity plays an important role in the attack model. The greater the fluctuation variance, the more underestimation of practical excess noise on both the communication sides, and the more insecure keys generated. This demonstrates the permeability of the modified LOIA with random fluctuations under DM CV-QKD protocols.
\begin{figure}[t]
\includegraphics[width=8.6cm]{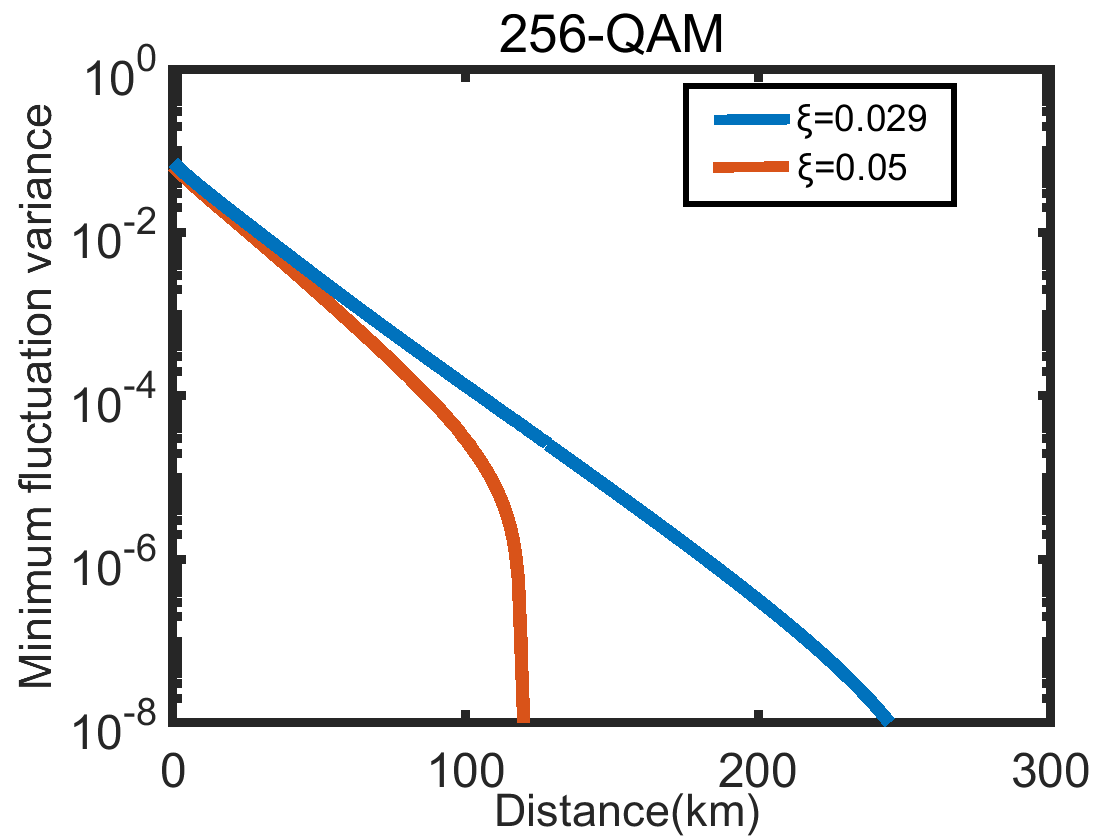}
\caption{\label{QAM_Vk}The minimal fluctuation variance of LO intensity for the 256-QAM-modulated protocol where Eve could acquire all keys without being detected. The modulation variance $V_A=6.332$, $\nu=0.039$ and $\beta=0.95$. }
\end{figure}

FIG. \ref{QAM_Vk} shows the minimum fluctuation variance when Eve could steal all the keys without being detected. We consider the case with the estimated channel excess noise $\xi_e=0.029$ and $\xi_e=0.05$, respectively. 
Eve could mask his attack to steal the secret keys by simply making the LO intensity fluctuate randomly with a very small fluctuation variance.

As can be seen from the above, within the DM CV-QKD system under the modified LOIA with random fluctuation, both Alice and Bob overestimate certain areas of the secret key rate so that Eve could hide her attacks on the signal pulse. 
The effect of the attack is closely related to the fluctuation variance of LO intensity. 
The greater the fluctuation variance, the more the communication parties overestimate the secret key rate.
As long as Eve chooses the appropriate fluctuation variance, she can obtain all the key information without being detected. The attack effect is more pronounced at high channel noise and long range.
\section{\label{sec:level6}Discussion}
\begin{figure}[t]
\includegraphics[width=8.6cm]{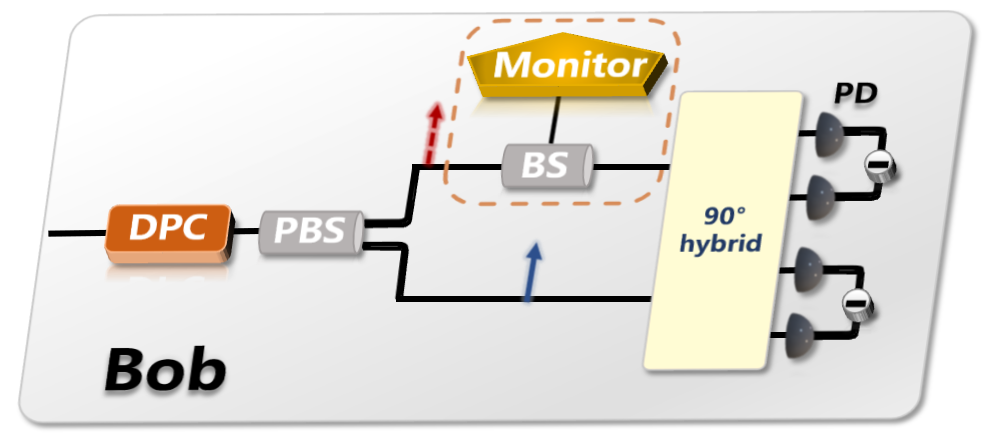}
\caption{\label{countermeasure}The practical setup of the real-time monitoring of the LO intensity. It consists of monitoring the intensity of each LO pulse by splitting a small part of the LO pulses and calibrating the instantaneous intensity. The dotted line diagram depicts the schematic diagram of the LO monitoring module.}
\end{figure}

In the practical environment, it has been analysed that the LO intensity level during key distribution process in the presence of Eve's manipulation cannot be determined directly using the initial calibrated LO intensity. Moreover, in the modified LOIA model with random fluctuations, the statistical mean of LO intensity is consistent with the calibrated LO intensity value. Traditional commonly-used defense countermeasures including monitoring the mean intensity and optical power of LO or monitoring shot-noise will fail. We must carefully monitor the instantaneous fluctuations in LO intensity and the corresponding countermeasure is to design modules that allow the real-time monitoring of LO intensity. FIG. \ref{countermeasure} gives a schematic setup of the countermeasure. It involves the LO intensity monitoring module which splits a small portion with an asymmetric splitter and monitors the instantaneous value of LO intensity. Then communication parties uses the monitored LO intensity to scale the environment measurements. If the intensity of each LO pulse is calibrated to obtain the outcomes during the key distribution, Eve's attack on LO can be avoided, making the practical DM CV-QKD system more robust. However, the countermeasure requires a higher degree of accuracy for the real-time monitoring technology. 
Moreover, we think that other schemes are also effective, such as the local local oscillator (LLO) CV-QKD system, the measurement-device-independent (MDI) QKD protocol and so on. But newly introduced security vulnerabilities and the difficulty of experimental implementation should also be carefully considered.

In previous LOIA model, Eve hides the excess noise introduced by her eavesdropping behaviour by overall attenuating the LO intensity. In this paper, we focus on the random fluctuating property of LO intensity and propose a modified LOIA model with random fluctuations with invariant mean value. 
The manipulation of LO intensity avoids detection by LO intensity and power monitoring technology or shot-noise monitoring technology.
The theoretical parameter estimation model constructed in practical DM CV-QKD system suggests that the underestimation of excess noise leads to key leakage without being detected by both communicating parties. From the simulation results, it is found that the fluctuation variance ${V_k}$ affects the excess noise and key rate estimated by both communicating parties greatly. If the fluctuating variance is relatively large, Eve can secretly steal all the keys, thus seriously compromising system security. 

Similarly, not only the transmitted local oscillator (TLO) CV-QKD system, the modified attack model also poses a serious challenge to the security of the pilot in the LLO CV-QKD system.
Traditionally, the communication parties use the mean intensity of the pilot to quantify the the trusted phase noise, and then calculate the system secret key rate.
Eve could exploit the loophole of the random fluctuations in pilot intensity to reduce the trusted phase noise \cite{shao2021phase, shao2022phase}, thereby hiding her attack on signal pulses.
\section{\label{sec:level6}Conclusion}
In conclusion, we propose a modified LO intensity attack with random fluctuations that can successfully mask Eve's eavesdropping behavior and evade commonly-used monitoring technologies. To analyze its attack effect, modified LO intensity attack is carried out to observe the parameter estimation of semidefinite program modeling against practical discrete-modulated CV-QKD system. The theoretical simulation results under the Gaussian channel show that even very small random fluctuations in modified LO intensity attack will cause no secure keys generate. The stealthy quantum hacking can also be launched in the practical LLO CV-QKD system with random fluctuations in pilot intensity.
Therefore, in a realistic CV-QKD system, the LO intensity should be well monitored and stabilised. The modified LO intensity attack model with random fluctuations in this paper places greater demands on current real-time intensity monitoring techniques. The related study will be of great significance to the practical security analysis of CV-QKD.
\section{\label{sec:level6}ACKNOWLEDGMENTS}
This research was supported by the National Natural Science Foundation of China (62001044), the Basic Research Program of China (JCKY2021210B059), the Equipment Advance Research Field Foundation (315067206), and the Fund of State Key Laboratory of Information Photonics and Optical Communications
(IPOC2021ZT02).
\appendix
\nocite{*}


\bibliography{apssamp.bib}
\end{document}